\documentclass[aps,floats,showpacs,preprint]{revtex4} 
\usepackage{graphicx}
\usepackage{epsfig}
\usepackage{psfig}
\usepackage{amsmath}
\usepackage{hyperref}

\def\PRD#1{{ Phys.\ Rev.} {\bf D#1}}
\def\NPB#1{{ Nucl.\ Phys.} {\bf B#1}}
\def\PLB#1{{Phys.\ Lett.} {\bf B#1}}

\def\be{\begin{equation}}
\def\ee{\end{equation}}
\def\bea{\begin{eqnarray}}
\def\eea{\end{eqnarray}}

{\newcommand{\lsim}{\mbox{\raisebox{-.6ex}{~$\stackrel{<}{\sim}$~}}}
{\newcommand{\gsim}{\mbox{\raisebox{-.6ex}{~$\stackrel{>}{\sim}$~}}}
\def\sss{\scriptscriptstyle}

\begin{document}

\title{\bf Dark matter and leptogenesis in gauged $B-L$ symmetric 
models embedding $\nu$MSM}

\author{Narendra Sahu$^a,b$}
\email{narendra@prl.res.in}
\author{Urjit A. Yajnik$^b$}
\email{yajnik@phy.iitb.ac.in}
\affiliation{$^a$Physical Research Laboratory, Ahmedabad-380 009, India}
\affiliation{$^b$Department of Physics, Indian Institute of Technology,
Bombay, Mumbai 400076 India}                   

\begin{abstract}
We study the phenomenon of baryogenesis via leptogenesis in the 
gauged $B-L$ symmetric models by embedding the currently proposed 
model $\nu MSM$. It is shown that the lightest right handed 
neutrino of mass $100 GeV$ satisfy the leptogenesis constraint 
and at the same time representing a candidate for the cold dark 
matter. We discuss our results in parallel to the predictions 
of $\nu MSM$.

\end{abstract}
\pacs{98.80.Cq, 14.60.St, 95.35.+d}
\maketitle
\section{Introduction}
At present the atmospheric neutrino data~\cite{atmos_data} 
in the $\nu_{\mu}-\nu_{\tau}$ oscillation and the solar neutrino 
data~\cite{solar_data} in the $\nu_e-\nu_{\mu}$ oscillation 
experiments are highly suggestive to believe small masses 
of the light neutrinos ($\leq$ 1eV), either Dirac or Majorana. 
Assuming that the neutrinos are of Majorana type the small masses 
can be understood through the seesaw mechanism~\cite{
seesaw_group}, which involves the right handed neutrinos into 
the electroweak model, invariant under all the gauge transformations. 

At the minimal cost we can add two right handed neutrinos to the
standard model ($SM$) Lagrangian to explain the tiny mass scales;
the atmospheric neutrino mass ($\Delta_{atm}=\sqrt{|m_3^2-m_2^2|}$) 
and the solar neutrino mass ($\Delta_{sun}=\sqrt{|m_2^2-m_1^2|}$). 
However, in this scenario the seesaw mechanism gives rise to one 
of the light neutrino mass to be exactly zero. This is unwelcome 
if the neutrino masses are partially degenerate, albeit the 
hierarchical mass spectrum of the light neutrinos can be 
conspired in this scenario. Since the exact
mass scales of the light neutrinos are not known yet, we therefore add
three right handed neutrinos, gauge invariantly, to the $SM$ Lagrangian.

In the thermal scenario the $CP$ violating decay of the right handed
Majorana neutrinos can potentially explain the matter antimatter
asymmetry~\cite{fukugita.86}, defined by
\be
\frac{n_B}{n_{\gamma}}=6.1\times 10^{-10},
\ee
of the present Universe as predicted by the Wilkinson Microwave
Anisotropy Probe (WMAP)\cite{spergel.03}. This requires the
scale of operation of right handed neutrinos to be $\gsim 10^8$
GeV~\cite{davidson&ibarra.02} and hence far beyond our 
hope to be verified in the near future accelerators. 

An alternative is to consider the mechanisms which work at TeV 
scale~\cite{sahu&yaj_prd.04,king&yan.04,chun.05}. In 
ref.~\cite{sahu&yaj_prd.04} it was proposed that the spontaneous 
breaking of the $B-L$ gauge symmetry gives rise to a raw lepton 
asymmetry. The preservation of lepton asymmetry then requires a 
limited wash out through the lepton violating interactions mediated 
by the right handed neutrinos and hence requiring the mass scale 
($M_1$) of lightest right handed neutrino ($N_1$) to be at the TeV scale 
or less. This needs to rethink whether these low mass scales of $N_1$ 
can be compatible with the seesaw mechanism to give rise the Majorana 
mass matrix of the light neutrinos to be 
\be
m_\nu=-m_D^T M_R^{-1}m_D.
\label{type-I-seesaw}
\ee
In particular, it was shown in ref.~\cite{sahu&yaj_prd.04} that a 
TeV mass of $N_1$ is compatible with the seesaw if we assume that 
the Dirac mass matrix of the neutrinos is two orders less than 
that of charged leptons mass matrix.

Recently ``$\nu MSM$" model has been proposed~\cite{shaposnikov.05}. 
In this model the right handed neutrinos are singlet under the $SM$ 
gauge group. The mass of $N_1$ in this case is constrained to
\be
2 KeV\leq M_1 \leq 5 KeV ,
\label{darkmatter-con}
\ee
where the lower bound comes from the Cosmic Microwave Background (CMB) 
and the matter power spectrum inferred from Lymen $\alpha$ forest 
data~\cite{viel&com.05} and the upper bound comes from the 
radiative decays of singlet right handed neutrinos in dark 
matter ($DM$) halos limited by X-ray observations~\cite{abazajian&com.01}.
This requires, from equations. (\ref{type-I-seesaw}) and 
(\ref{darkmatter-con}), that the Dirac Yukawa coupling $h_{\nu}\sim 10^{-10}$ 
for $m_\nu$=0.1 eV. The tiny Yukawa coupling in this scenario makes 
the lightest right handed neutrino decoupled from the thermal bath 
through out it's evolution. However, this constraint is not applicable 
to $N_2$ and $N_3$, the second and third generation of right handed 
neutrinos. Hence they come into equilibrium through the large Yukawa 
couplings. This permits the authors in ref.~\cite{smirnov_prl.98} to 
consider a mechanism to create a net lepton asymmetry in the right 
handed neutrino sector through oscillations~\cite{smirnov_prl.98}. 
The net lepton asymmetry created in the right handed sector is then 
transferred to the left handed sector through the Yukawa coupling 
\be
\mathcal{L}_{Y}=(h_{\nu})_{ij}\phi\bar{\psi}_{Li} N_{j}.
\label{yukawa}
\ee
The lepton asymmetry is then transferred to baryon asymmetry 
through the nonperturbative sphaleron processes~\cite{krs.85}. 

An important issue of the $\nu MSM$ model is that the mass of 
$N_1$ is severely constrained from the hot $DM$ consideration. 
Further the lepton asymmetry produced by any mechanism other 
than the sterile neutrino oscillation will continue to survive 
and hence invaliding the leptogenesis constraints on the right 
handed neutrino masses. Moreover the Dirac Yukawa couplings are 
very tiny. 
     
As an alternative, in the present case we consider the low energy 
left-right symmetric model~\cite{leftright_group}. Since $B-L$ is 
a gauge symmetry of the model any primordial asymmetry is erased. 
Further advantage of considering this model is that it can be 
easily embedded in the unified models like $SO(10)$ or Pati-Salam 
and at the same time it can embed the ``$\nu MSM$" by gauging the 
$B-L$ symmetry. In this model, by assuming a normal mass hierarchy 
in the right handed neutrino sector, we discuss the role of lightest 
right handed neutrino in leptogenesis as well as for dark matter. It 
is shown that the lightest right handed neutrino of mass $100 GeV$ can 
be a candidate for $DM$ as well as satisfying the erasure constraint 
required for the preservation of lepton asymmetry.

Rest of the manuscript is arranged as follows. In sec-II we 
discuss the leptogenesis in the left-right symmetric models and 
then elucidate the possibility of bringing down the mass scale 
of $N_1$ to TeV scale or less. In sec-III the constraint on the 
$B-L$ breaking scale is discussed. In sec-IV we discuss the 
constraint on the mass scale of lightest right handed neutrino from 
the Flavor Changing Neutral Current (FCNC). Sec-V is devoted to discuss 
the possibility of TeV scale right handed neutrinos for the candidate 
of dark matter. Finally in sec-VI we summarize our results and put the 
conclusions.  

\section{Leptogenesis in gauged $B-L$ symmetric models and the 
possibility of TeV scale right handed neutrino}
In the following we consider the left-right symmetric model where 
$B-L$ gauge symmetry emerges naturally. However, the arguments to 
be advocated below will remain valid as long as $B-L$ is a gauge 
symmetry of the model.

\subsection{Spontaneous CP-violation in L-R symmetric model and 
leptogenesis}
The main attraction of the left-right symmetric model lies in 
the lepton sector. The right handed neutrinos, which were singlet 
under the SM gauge group $SU(2)_L\times U(1)_Y$, now non-trivially 
transforms under the left-right symmetric gauge group 
$SU(2)_L\times SU(2)_R\times U(1)_{B-L}$. Since the right 
handed neutrinos possess the $B-L$ quantum number by one unit 
the Majorana mass can violate lepton number by two units and 
hence is a natural source of lepton asymmetry in the model. 

The Higgs sector of the left-right symmetric model is very rich. 
It consists of two scalar triplets $\Delta_L$ and $\Delta_R$ which give 
Majorana masses to the right handed neutrinos and a bidoublet 
$\Phi$ which gives Dirac masses to the charged leptons and quarks.  
We assume that all the Yukawa couplings in the Higgs 
potential of the model are real. Thus the Lagrangian respects 
CP symmetry. The complex nature of the neutrino masses then come 
through the VEVs of the neutral Higgses~\cite{scpv-group}. In general 
there are four neutral Higgses in the model can potentially acquire 
VEVs and thereby breaking the left-right symmetry down to $U(1)_{em}$. 
Hence there are four phases associated with the neutral Higgses. 
However, the remnant global symmetry  $U(1)_L\times U(1)_R$ allows 
us to set two of the phases to zero. Therefore, only two of the phases 
have physical significance. 
    
The breaking of left-right discrete symmetry in the early Universe 
gives rise to domain walls. It was shown in~\cite{cline&yajnik_prd.02} 
that within the thickness of the domain walls the net $CP$ violating 
phase becomes position dependent. Under these circumstances the 
preferred scattering of $\nu_L$ over its CP-conjugate state ($\nu_L^c$) 
produce a net raw lepton asymmetry~\cite{
cline&yajnik_prd.02}
\be
\eta^{\rm raw}_{\sss L} \cong 0.01\,  v_w {1\over g_*}\,
        {M_1^4\over T^5\Delta_w}
\label{eq:ans2}
\ee
where $\eta^{\rm raw}_{\sss L}$ is the ratio of $n_L$ to the entropy 
density $s$. In the right hand side $\Delta_w$ is the wall width and 
$g_*$ is the effective thermodynamic degrees of freedom at the epoch 
with temperature $T$.  Using $M_1 =f_1 \Delta_{\sss T}$,
with $\Delta_{\sss T}$ is the temperature dependent VEV acquired by 
the $\Delta_{\sss R}$ in the phase of interest, and 
$\Delta_w^{-1} = \sqrt{\lambda_{eff}}\Delta_{\sss T}$ in equation 
(\ref{eq:ans2}) we get
\be
\eta^{\rm raw}_L \cong 10^{-4} v_w
\left(\frac{\Delta_{\sss T}}{T} \right)^5 f_1^4
\sqrt{\lambda_{eff}}.
\ee
Here we have used $g_*=110$. Therefore, depending on the various
dimensionless couplings, the raw asymmetry may lie in the range 
$O(10^{-4}~-~10^{-10})$. 

\subsection{TeV scale right handed neutrino and lepton asymmetry}
In the previous section we saw that a net raw lepton asymmetry 
($\eta^{\rm raw}_L$) is generated through the scattering of light 
neutrinos on the domain wall. However, it may not be the final 
asymmetry. This is because of the thermally equilibrated lepton 
violating processes mediated by the right handed neutrinos can 
erase the produced asymmetry. Therefore, a final asymmetry and hence 
the bound on right handed neutrino masses can only be obtained by 
solving the Boltzmann equations~\cite{Boltzmann_equations}. We 
assume a normal mass hierarchy in the 
right handed neutrino sector. In this scenario, as the temperature 
falls, first $N_3$ and $N_2$ go out of thermal equilibrium while 
$N_1$ is in thermal equilibrium. Therefore, it is the density and 
mass of $N_1$ are important in the present case which enter into 
the Boltzmann equations. The relevant Boltzmann equations for the 
present purpose are~\cite{sahu&yaj_prd.04}
\bea
\frac{dY_{N1}}{dZ} &=& -(D+S)\left(Y_{N1}-Y^{eq}_{N1}\right)
\label{boltzmann.1}\\
\frac{dY_{B-L}}{dZ} &=& -W Y_{B-L}
\label{boltzmann.2},
\eea
where $Y_{N_1}$ is the density of $N_1$ in a comoving volume and that 
of $Y_{B-L}$ is the $B-L$ asymmetry. The parameter $Z=M_1/T$. The 
various terms $D$,$S$ and $W$ are representing the decay, scatterings 
and the wash out 
processes involving the right handed neutrinos. In particular, 
$D=\Gamma_D/ZH$, with 
\be
\Gamma_D=\frac{1}{16 \pi v^2}\tilde{m}_1 M_1^2,
\label{decay}
\ee
where $\tilde{m}_1=(m_D^{\dagger}m_D)_{11}/M_1$ is called the effective 
neutrino mass parameter. Similarly $S=\Gamma_S/HZ$ and $W=\Gamma_W/HZ$. Here 
$\Gamma_S$ and $\Gamma_W$ receives the contribution from $\Delta_L=1$ 
and $\Delta_L=2$ lepton violating scattering processes. The dependence 
of the scattering rates involved in $\Delta_{\rm L}=1$ lepton violating 
processes on the parameters $\tilde{m}_1$ and $M_1$ is similar to that 
of the decay rate $\Gamma_{D}$. As the Universe expands these $\Gamma$'s 
compete with the Hubble expansion parameter. Therefore in a comoving 
volume we have
\be
\left(\frac{\gamma_{D}}{sH(M_1)}\right), \left(\frac{
\gamma^{N1}_{\phi,s}}{sH(M_1)}\right), \left(\frac{
\gamma^{N1}_{\phi,t}}{sH(M_1)}\right) \propto k_1\tilde{m}_1.
\label{dilution}
\ee
On the other hand, the dependence of the $\gamma$'s in
$\Delta_{\rm L}=2$ lepton number violating processes on
$\tilde{m}_1$ and $M_1$ are given by
\be
\left(\frac{\gamma^l_{N1}}{sH(M_1)}\right), \left(\frac{
\gamma^l_{N1,t}}{sH(M_1)}\right) \propto k_2 \tilde{m}_1^2 M_1.
\label{washout}
\ee
Finally there are also lepton conserving processes
where the dependence is given by
\be
\left(\frac{\gamma_{Z'}}{sH(M_1)}\right) \propto k_3 M_1^{-1}.
\label{l-conserve}
\ee
In the above equations (\ref{dilution}), (\ref{washout}), 
(\ref{l-conserve}), $k_i$, $i=1,2,3$ are dimensionful constants 
determined from other parameters. Since the lepton conserving 
processes are inversely proportional to the mass scale of $N_1$, 
they rapidly bring the species $N_1$ into thermal equilibrium 
for $T\gg M_1$. Further smaller the values of $M_1$, the 
washout effects (\ref{washout}) are negligible because of their 
linear dependence on $M_1$. This is the regime in which we are
while solving the Boltzmann equations in the following.

The equations (\ref{boltzmann.1}) and (\ref{boltzmann.2}) are
solved numerically. The initial $B-L$ asymmetry is the net raw
asymmetry produced during the $B-L$ symmetry breaking phase
transition by any thermal or non-thermal processes. As such we
impose the following initial conditions
\be
Y^{in}_{N1}=Y^{eq}_{N1}~~ {\mathrm and}~~ Y^{in}_{B-L}=\eta^{raw}_{B-L}, 
\label{initial-cond}
\ee
by assuming that there are no other processes creating lepton 
asymmetry below the $B-L$ symmetry breaking scale. This requires 
$\Gamma_D\leq H$ at an epoch $T\geq M_1$ and hence lead to a 
bound~\cite{fglp_bound}
\be
m_\nu < m_* \equiv 4\pi g_*^{1/2}\frac{G_N^{1/2}}{\sqrt{2}G_F}
= 6.5\times10^{-4}eV.
\label{eff_par_cons}
\ee  
Alternatively in terms of Yukawa couplings this bound reads 
\be
h_{\nu}\leq 10 x, ~~~{\rm with}~~ x=(M_1/M_{pl})^{1/2}. 
\label{yukawa_const}
\ee
\begin{figure}[ht]
\begin{center}
\epsfig{file=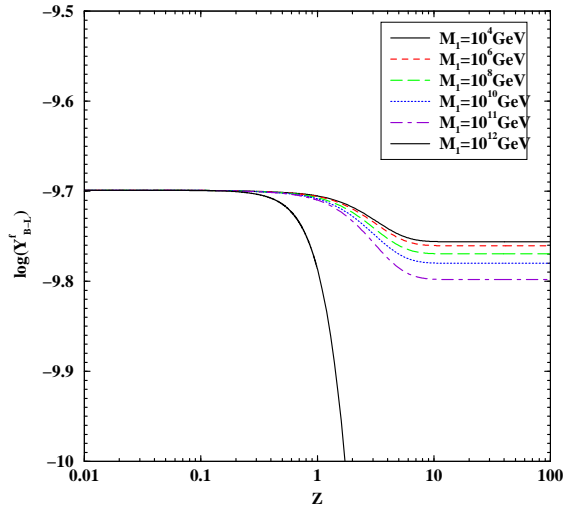, width=0.45\textwidth}
\caption{The evolution of B-L asymmetry for different values of
$M_1$ shown against $Z(=M_1/T)$ for $\tilde{m}_1=10^{-4}$eV
and $\eta^{raw}_{B-L}=2.0\times 10^{-10}$}
\label{figure-1}
\end{center}
\end{figure}

At any temperature $T\geq M_1$, wash out processes involving $N_1$ are
kept under check due to the $\tilde{m}_1^2$ dependence in (\ref{washout})
for small values of $\tilde{m}_1$. As a result a given
raw asymmetry suffers limited erasure. As the temperature
falls below the mass scale of $N_1$ the wash out processes
become negligible leaving behind a final lepton asymmetry.
Fig.\ref{figure-1} shows the result of solving the Boltzmann 
equations for different values of $M_1$. An important conclusion 
comes from this figure is that for smaller values of $M_1$ the wash out 
effects are tiny. Hence by demanding the initial raw asymmetry is 
the required asymmetry of the present Universe we can conspire the 
mass scale of $N_1$ to be as low as 1 TeV. For this value 
of $M_1$, using equation (\ref{yukawa_const}), we get the constraint 
$h_{\nu}\leq 10^{-7}$. Further lowering of $M_1$ needs $h_{\nu}<10^{-7}$. 
This was the prediction of the model $\nu MSM$ to keep $M_1$ in KeV 
range, albeit the leptogenesis mechanism was different. However in 
the present case as we see in section-IV, the bound on $M_1$ is very 
much tight from the flavor changing neutral current unless we allow 
sufficiently small Yukawa couplings.         

Note that in equation (\ref{boltzmann.2}) we assume that there are 
no other sources producing lepton asymmetry below the $B-L$ symmetry 
breaking phase transition. This can be justified by considering 
small values of $h_{\nu}$, since the $CP$ asymmetry parameter $\epsilon_1$ 
depends quadratically on $h_{\nu}$. Hence for $h_{\nu}\leq 10^{-7}$ 
the lepton asymmetry $Y_L \leq O(10^{-14})$, which is far less 
than the raw asymmetry produced by the scatterings of neutrinos on the 
domain walls. This explains the absence of lepton asymmetry 
generating term in equation (\ref{boltzmann.2}).

\section{Constraint on the B-L breaking scale}
Below the mass scale of $N_1$ the lepton conserving processes 
$N_1 N_1\rightarrow f\bar{f}$ mediated by the $Z'$ boson fall 
out of equilibrium. Here $f$ and $\bar{f}$ are the $SM$ 
fermions and anti fermions. The cross-section is given 
as
\be
\sigma (N_1 N_1\rightarrow \sum_f f\bar{f})\sim \frac{1}{4\pi}
\frac{E^2}{v_{B-L}^4},
\label{z'-cross}
\ee
where we have used the mass of $Z'$ boson $M_{Z'}=g' v_{\rm B-L}$, with 
$v_{\rm B-L}$ is the $B-L$ symmetry breaking scale. At the epoch 
$T\sim M_1$ the rate of lepton conserving process mediated by 
the $Z'$ boson is given by 
\be
\Gamma_{Z'} = n_{N_1}<\sigma v>,
\label{cross-rate}
\ee
where $n_{N_1}$ is the density of $N_1$ at that epoch. Further at 
the epoch $T\gsim M_1$, $n_{N_1}= n_{N_1}^{eq} =2T^3/\pi^2$. 
Hence substituting it in equation (\ref{cross-rate}) and using $\sigma$ 
from equation (\ref{z'-cross}) we get 
\be
\Gamma_{Z'} = \frac{1}{2\pi^3}\frac{M_1^5}{v_{\rm B-L}^4}.
\label{cross-rate-1}
\ee
Requiring $\Gamma_{Z'}\leq H(M_1)$ we get 
\be
v_{\rm B-L}\geq \left(\frac{M_{pl}}{2\pi^3\times 1.67 g_*^{1/2}}\right)^{1/4} 
M_1^{3/4}\sim 10^6 GeV\left( \frac{M_1}{100 GeV}\right)^{3/4}.
\label{B-L const}
\ee 
This tells us that for $M_1=100 GeV$, the $B-L$ breaking scale is 
greater than $10^6$ GeV. This is in well agreement with equation 
(\ref{yukawa_const}) for $h_{\nu}\leq 10^{-7}$.

\section{FCNC constraint on the mass scale of $N_1$}
In a flavor basis the Lagrangian describing the neutral current 
for one generation of fermions is given as 
\be
\mathcal{L}\simeq \frac{g}{2 \cos \theta_W}Z^{\mu}\bar{\nu}_{eL}
\gamma_{\mu}\nu_{eL},
\label{neu-current}
\ee
where $\theta_W$ is the weak mixing angle. Rewriting equation 
(\ref{neu-current}) in a mass basis we get 
\be
\mathcal{L}\simeq \frac{g}{2\cos \theta_W}Z^{\mu}\left[\cos^2\theta
\bar{\nu}_1\gamma_{\mu}L\nu_1+\sin^2 \theta
\bar{N_1}\gamma_{\mu}L N_1+\cos\theta \sin\theta(\bar{N_1}
\gamma_{\mu}L\nu+\bar{\nu_1}\gamma_{\mu}L N_1)\right],
\label{fcnc-mass-current}
\ee
where $L$ is the left-handed projection operator and $\theta$ is 
the mixing angle and is given by 
\be
\theta=\frac{m_D}{M_1}=\left(\frac{m_{\nu}}{M_1}\right)^{1/2},
\label{theta}
\ee
where we have used the equation (\ref{type-I-seesaw}). Thus there is a 
flavor changing neutral current in the model as given by the third term 
in equation (\ref{theta}). This is unlike the case 
in $SM$. Hence by requiring $\theta$ to be small, the flavor changing 
neutral current can be suppressed. Using the current bound $m_{\nu_e}
\leq 0.6 eV$ from the neutrino less double beta decay experiment
~\cite{klapdor} we get from equation (\ref{theta}) that 
$\theta\lsim 10^{-6}$ for $M_1\gsim 1 TeV$. 

On the other hand, if we relax the upper bound on $\theta$ by three 
orders larger than the above bound then we get a lower bound on 
$M_1$ to be $\gsim 1 GeV$. This will allow the following decay width 
$\Gamma(Z\rightarrow \nu N)\sim \theta^2 165 MeV$~\cite{mohapatra-pal-book}, 
for mass of $N_1$ ranging from 1 GeV to 80 GeV. If $\theta$ is 
large this decay has a distinctive signature through the decay 
modes of $N_1$. 
In particular, $\Gamma(N_1 \rightarrow 3\nu)\propto \theta^2$. 
Therefore, the above decay mode of Z boson is highly restricted.
 
Now we study the bound on $\theta$ by considering the magnitude of 
Dirac Yukawa coupling of the neutrinos. Since $\theta=m_D/M_1=
h_{\nu} v/M_1$, we can achieve small values of $\theta$ by demanding 
$h_\nu\ll h_e$ even for small values of $M_1$. This was the prediction 
of $\nu MSM$ model, where the Yukawa coupling $h_{\nu}$ was 
required to be very small.
     
\section{Dark matter constraint on mass scale of $N_1$}
One of the important questions in cosmology is that how much the 
masses of the galaxies contribute to the critical density 
\be
\rho_c=\frac{3H_0^2}{8\pi G_N}\equiv 10^4 h^2 eV/cm^{3}
\label{critical-density}
\ee
of the present Universe. Here $H_0=h\times 100 Km s^{-1}Mpc^{-1}$, 
with $0.4\lsim h\lsim 1.0$ is the Hubble expansion parameter that 
is observed today. The best fit value, combinely given by the WMAP, 
2dFGRS and Lymen $\alpha$ forest data, is $h=0.72\pm 0.03$~
\cite{spergel.03}. On the other hand, the $\Omega$ parameter defined 
for the total density of the Universe is given by
\be
\Omega_{tot}=\Omega_m+\Omega_{\Lambda}=1.02\pm 0.02,
\label{density-unv}
\ee
where the various $\Omega$'s are defined as $\Omega_i=(\rho_i/\rho_c)$. 
Equation (\ref{density-unv}) indicates that the present Universe 
is flat with the mass density contributed by the galaxies is 
approximately equal to it's critical density. The best fit value 
for the present matter component of the Universe, combinly given 
by the WMAP with 2dFGRS and Lymen alpha forest data, is 
$\Omega_{m}=0.133\pm 0.006/h^2$. However, the baryonic component of 
matter is found to be $\Omega_B=0.0226\pm 0.0008/h^2$. This implies 
that the present Universe contains significant amount of 
non-baryonic matter which is given by $\Omega_{NB}=0.1104/h^2$. 
The missing matters are usually treated as dark matter ($DM$). 

An important issue of the particle physics and cosmology 
is that the nature of dark matter and its role in the evolution 
of the Universe. Had it been the cold dark matter it had played 
an important role in the formation of large scale structure 
of the Universe. At present the contribution of light neutrinos 
having masses varying from $5\times 10^{-4}$ eV to 1 MeV 
is~\cite{pdg.05} 
\be
\Omega_{\nu}\leq 0.0076/h^2~~~~~~~~95\% ~~C.L.
\ee
However, this is not sufficient to explain the non-baryonic 
component of matter. In the present model we propose that the 
lightest right handed neutrino can be a suitable candidate for 
cold $DM$ for which the life time of $N_1$ must satisfy the constraint, 
$\tau_{N_1} > 2t_0$,     
where $t_0$ is the present age of the Universe. Alternatively we require 
$\Gamma_{N_1}<{H_0}$, the present Hubble expansion parameter.
This gives the constraint on the Dirac mass of the neutrino to be 
\be
(m_D^{\dagger}m_D)_{11}<1.19\times 10^{-40}GeV^2 
\left(\frac{10^3GeV}{M_1}\right).
\label{Dirac-mass-con}
\ee
A similar constraint on the Dirac mass of the neutrino was obtained 
in ref.~\cite{babu&moh_plb.89} for $N_1$ to be a candidate of cold 
$DM$. 

Since the massive neutrinos are stable in the cosmological time 
scale we have to make sure that it should not over-close the Universe. 
For this we have to calculate the density of the heavy neutrino $N_1$ 
at the present epoch of temperature $T_0=2.75^\circ K$. The number 
density of $N_1$ at present is given by 
\be
n_{N_1}(T_0)=n_{N_1}(T_D)\left(\frac{T_0}{T_D}\right)^3
\label{neu_abun}
\ee
where $T_D$ is the temperature of the thermal bath when the massive 
neutrinos got decoupled. This can be calculated by considering the  
out of equilibrium of the annihilation rate $\Gamma_{ann}$ of the 
process $N_1 N_1\rightarrow f\bar{f}$. We assume that at a 
temperature $T_D$
\be
\Gamma_{ann}/H(T_D)\simeq 1,
\label{dec-era}
\ee
where $\Gamma_{ann}$ is essentially given by equation (\ref{cross-rate}) 
and 
\be
H(T_D)=1.67g_*^{1/2}\frac{T_D^2}{M_{Pl}}
\label{hubble}
\ee
is the Hubble expansion parameter during the decoupled era. 
Considering the effective four-Fermi interaction of the annihilation 
processes $\sigma$ can be parameterized as~\cite{babu&moh_plb.89}
\be
\sigma_{N_1}=\frac{G_F^2M_1^2}{2\pi}c,
\label{crossection}
\ee
where $c$ is the compensation factor and is taken 
to be $O(10^{-2})$. Further $n_{N_1}$ is the density of $N_1$ at 
an epoch $T\sim M_1$. At any temperature $T$, the density 
distribution $n_{N_1}$ is given by 
\be
n_{N_1} (T) =2\left(\frac{M_1T}{2\pi}\right)^{3/2} exp \left( 
-\frac{M_1}{T}\right).
\label{eqb-value}
\ee
Using (\ref{hubble}), (\ref{crossection}) and (\ref{eqb-value}) 
in equation (\ref{dec-era}) we get 
\be
\frac{\Gamma_{ann}}{H(T_D)}=1.2\times 10^{-2} g_*^{-1/2}N_{ann}
G_F^2 M_1^3 M_{Pl}c z_D^{1/2} exp(-z_D)\simeq 1,
\label{decay-rate}
\ee
where $z_D=M_1/T_D$ and $N_{ann}$ is the number of annihilation 
channels which we take $\approx 10$. Solving for $z_D$ from 
equation (\ref{decay-rate}) we get  
\be
z_D\approx \ln \left[\frac{N_{ann}}{82g_*^{1/2}}
\left(G_F^2 c M_1^3M_{Pl}\right)\right].
\label{z-decouple}
\ee
Using (\ref{decay-rate}) in equation (\ref{neu_abun}) we get 
\bea
n_{N_1}(T_0) &=&\frac{2}{(2\pi)^{3/2}}z_D^{3/2}exp(-z_D)T_0^3\nonumber\\
&=&\frac{2.016\times 10^{-11}}{cm^3}\left(\frac{TeV}{M_1}\right)^3
\left[1+0.02+0.21\ln\frac{M_1}{1TeV}\right].
\label{neu-density}
\eea
Now we can define the energy density of $N_1$ at the present epoch 
as 
\bea
\rho_{N_1}&=&n_{N_1}M_1\nonumber\\
&=&\frac{20.16}{cm^3}\left(\frac{1TeV}{M_1}\right)^2(1+correction).
\label{energy-density}
\eea
Using equations (\ref{critical-density}) and (\ref{energy-density}) we 
can get the $\Omega$ parameter for $N_1$ as
\bea
\Omega_{N_1} &=&\frac{\rho_{N_1}}{\rho_c}\nonumber\\
&=& \left(\frac{0.2016\times 10^{-2}}{h^2}\right) \left(
\frac{1TeV}{M_1}\right)^2.
\label{omega_n1}
\eea
Thus equation (\ref{omega_n1}) shows that for $M_1=1 TeV$ the 
contribution of $N_1$ to the present $DM$, $\Omega_{DM}=(0.1104/h^2)$ 
is two orders less. On the other hand, if we allow $M_1\simeq 100 
GeV$~\cite{dm-bound} then we can 
satisfy the present $DM$ constraint $\Omega_{DM}=(0.1104/h^2)$. In 
this mass limit of $N_1$ we get from equation 
(\ref{theta}) that the mixing angle $\theta\simeq 10^{-5}$. 

\section{Summary and Conclusion}
We studied the dark matter and leptogenesis constraints on 
the mass scale of lightest right handed neutrino in a gauged 
$B-L$ symmetric model. In this model the break down of the $B-L$ 
gauge symmetry produces a net raw lepton asymmetry which under 
goes a limited erasure for $\tilde{m}_1\lsim 10^{-4}eV$ and 
$M_1<10^{12}GeV$ and hence leaves behind the required lepton 
asymmetry which gets converted to the baryon asymmetry that is 
observed today. Therefore the assumption of raw lepton asymmetry 
of $\sim O(10^{-10})$ allows the mass scale of lightest right 
handed neutrino to be $1 TeV$ or less. However, for $M_1=1 TeV$ 
the contribution of $N_1$ to wards cold $DM$ is two orders less than 
the required value. On the other hand by requiring the mass scale of 
lightest right handed neutrino to be $O(10^2)$ GeV we can 
satisfy both leptogenesis as well as cold $DM$ constraint. Further 
in the left-right symmetric model for $M_1=100 GeV$ the 
mixing angle $\theta\lsim 10^{-5}$ and hence the flavor changing 
neutral current is suppressed. 

\section*{Acknowledgment}
NS wishes to thank Prof. S Uma Sankar and Prof. Utpal Sarkar for 
discussions and useful suggestions.


\end{document}